\documentclass{ckm}                 

\confname{Workshop on the CKM Unitarity Triangle, IPPP Durham, April
  2003}

\newcommand{\dkpi}{\ensuremath{D^{\pm} K^0_s \pi^{\mp}}}
\newcommand{\bdkpi}{\ensuremath{B^0 \to D^{\pm} K^0_s \pi^{\mp}}}

\newcommand{\Bbar}{\ensuremath{\overline{B} \rule{0ex}{1ex}^0}}

\newcommand{\ks}{\ensuremath{K^0_s}}

\newcommand{\fb}{\ensuremath{\mbox{fb}^{-1}}}

\def\bc{{_{\cal C}}}
\def\bu{{_{\cal U}}}
\def\sex{S_{\pi/2}}
\def\ssign{S_{-}}
\def\spi{S_\pi}

\def\bbartocbar{\overline{b}\rightarrow\overline{c}u\overline{s}}
\def\bbartoubar{\overline{b}\rightarrow\overline{u}c\overline{s}}

\def\bzbtodpm{\Bbar \rightarrow D^\pm \ks \pi^\mp}
\def\bztodmp{B^0 \rightarrow D^\mp \ks \pi^\pm}
\def\bztodm{B^0 \rightarrow D^- \ks \pi^+}
\def\bztodp{B^0 \rightarrow D^+ \ks \pi^-}
\def\bzbtodm{\Bbar \rightarrow D^- \ks \pi^+}
\def\bzbtodp{\Bbar \rightarrow D^+ \ks \pi^-}
\def\dmkspip{D^- K_s \pi^+}

\def\btodkpiall{$ B^0 \rightarrow D^{(*)\mp} K^{(*)0} \pi/\rho^\pm $}

\title{Measuring $2\beta + \gamma$ with Color-Allowed \bdkpi\ Decays}

\author{R.\ Aleksan\addressmark{a}, T.C.\ Petersen\addressmark{b}}
\address[a]{DAPNIA/SPP, Saclay, Gif-sur-Yvette, 91191 France}
\address[b]{LAL Bat.\ 208, Orsay BP 34, 91898 France}

\begin{document}

\begin{abstract}
We present a method to measure the weak phase $2\beta+\gamma$ in the
three-body decay of neutral $B^0(\Bbar)$ mesons to the final states
$\dkpi$. These decays are mediated by interfering amplitudes which are
color-allowed and hence relatively large. As a result, large CP violation
effects can be observed with high statistical significance. In addition, the
three-body decay helps to resolve discrete ambiguities present in measurements
of the weak phase. The experimental implications of conducting these
measurements are discussed, and the sensitivity of the method is evaluated
using a simulation.
\end{abstract}

\maketitle

\section{Introduction}
\vspace*{-1ex}
The violation of the CP symmetry is now established in the B meson sector and
the parameter $\sin(2\beta)$ is measured with precision by
BaBar~\cite{ref:babar-sin2b} and Belle~\cite{ref:belle-sin2b}.
However, the measurement of the other angles of the unitarity triangle are
necessary for a more comprehensive study of CP violation. These measurements
suffer from several difficulties related to the uncertainties of the theory,
the limited statistics and/or ambibuities in their extraction.  We have
recently proposed a new method~\cite{ref:APS} involving three-body B meson
decays, which circumvent these problems and allows to determine the angle
$\gamma = \arg{\left(-V_{ud} V_{ub}^* / V_{cd} V_{cb}^* \right)}$.

In this paper we extend this method to the weak phase
$\arg{\left(V_{td}^{* 2}V_{tb}^{2} V_{ub}^* V_{us}/ V_{cd}^{* 2}
V_{cb}V_{cs}\right)}$, which can be identified to the combination $2\beta +
\gamma$ of the unitarity triangle angles~\cite{ref:AKL}. Important
constraints on the theory will be obtained from the measurement of this phase.
In the following, we describe a new {\it time dependent} method involving the
entire area of the Dalitz plot of the 3-body decays \bdkpi.\\
The method involves no reconstruction into $CP$ eigenstates and avoids
problems with interfering DCSD.

\section{Color-Allowed \boldmath $B^0 \!\!\to\!\! \dkpi$ decays}
\vspace*{-1ex}
\label{sec:method-infinite-stat}
We investigate a way to circumvent the color suppression penalty by using
$B^0(\Bbar)$ decay modes which potentially offers significantly large
branching fractions and CP asymmetries. The particular decays
considered here are of the type \btodkpiall. These three body
decays may be obtained by popping a $q\bar{q}$ pair in color
allowed decays. Although modes where one or more of the three final state
particles is a vector can also be used, for clarity and simplicity only the
mode \bdkpi\ is discussed here.

\begin{figure}[htbp!]
\hbox to\hsize{\hss
\includegraphics[width=\hsize]{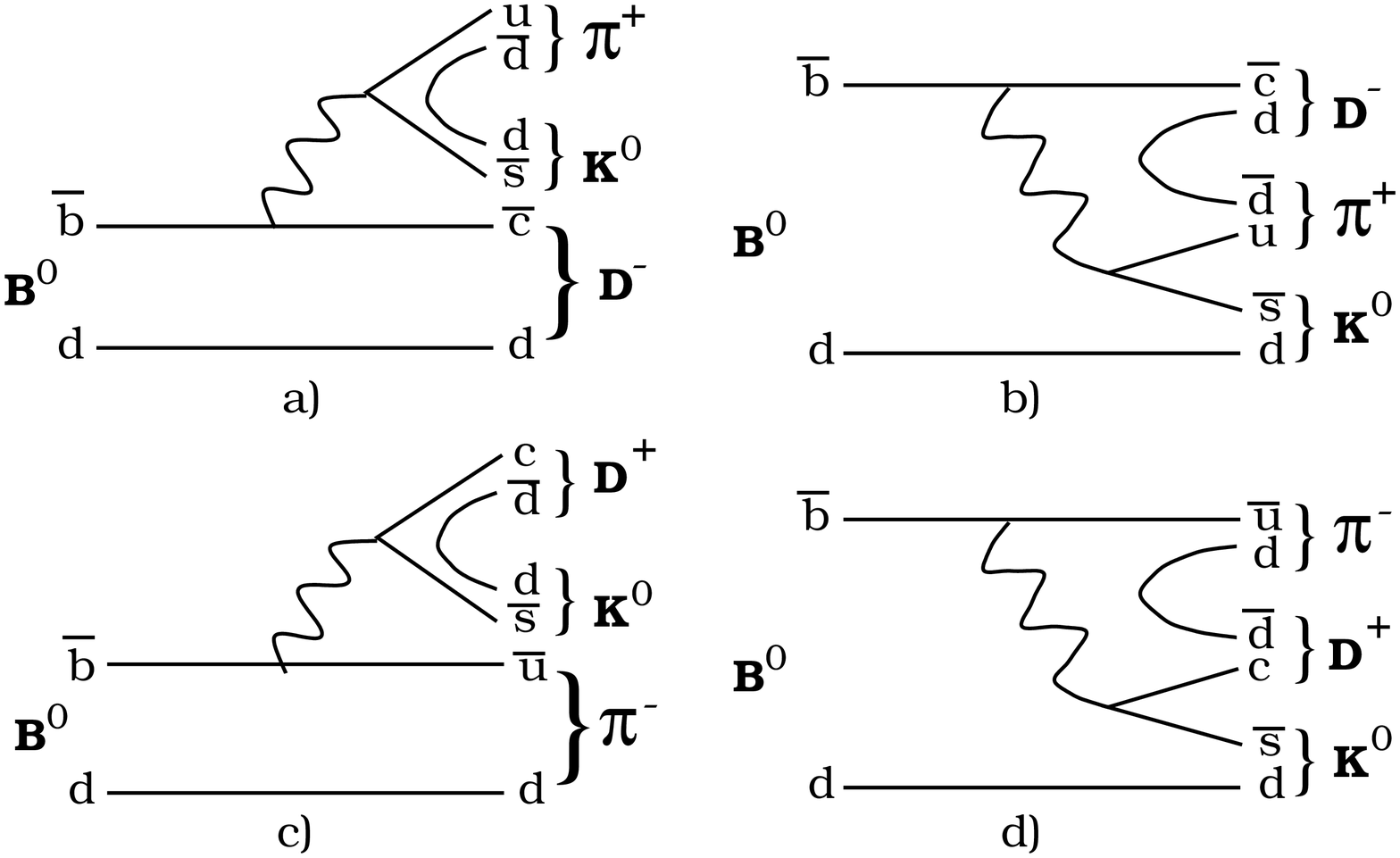}
\hss}
\caption{Feynman Diagrams for the decay $\bdkpi$ involving the CKM matrix
         element product $V_{cb}^*V_{us}$ or $V_{ub}^*V_{cs}$.}
\label{fig:diagrams}
\end{figure}

The diagrams leading to the final states of interest are shown in
Fig.~\ref{fig:diagrams}. As can be seen, the diagrams
(Fig.~\ref{fig:diagrams}a and~\ref{fig:diagrams}c) are both color-allowed and
of order $\lambda^3=\sin^3(\theta_c)$ in the Wolfenstein
parameterization~\cite{LW}, where $\theta_c$ is the Cabibbo mixing angle.

The general formalism describing how the weak angle is extracted from
observation of the time-dependent CP violation has been first presented in
\cite{ADKL}. However, in the present analysis additional information is
used to resolve the inherent ambiguities related to the method.

Selecting a particular point $i$ in the Dalitz plot, let us write the
amplitudes corresponding to the transitions in Fig.~\ref{fig:diagrams}a and b
on the one hand and~\ref{fig:diagrams}c and d on the other hand as
\vspace*{-1ex}
\begin{eqnarray}
{\cal A}_i(\bztodm) &=& A\bc_{i}e^{i\delta\bc_{i}}, \\
{\cal A}_i(\bztodp ) &=& A\bu_{i}e^{i\delta\bu_{i}} e^{i\gamma},
\label{eq:bzamp}
\end{eqnarray}
and their CP conjugates
\begin{eqnarray}
{\cal A}_i(\bzbtodp) &=& A\bc_{i}e^{i\delta\bc_{i}}, \\
{\cal A}_i(\bzbtodm ) &=& A\bu_{i}e^{i\delta\bu_{i}} e^{-i\gamma}, 
\label{eq:bzbamp}
\end{eqnarray}
where $\gamma$ is the relative phase of the CKM matrix elements
involved in this decay, and $A\bc$ ($A\bu$) and $\delta\bc$
($\delta\bu$) are the real amplitudes and CP-conserving strong
interaction phases.\\[-0.8ex]
The time dependent probability density function (PDF) of the transition
$B^0 \hspace*{-3.0ex} \raisebox{2.2ex}{\tiny (-- \hspace*{-3.1ex} --)}
\rightarrow \dmkspip $ is
\begin{equation}
\begin{array}{l}
\displaystyle {\cal P}r \left( 
B^0 \hspace*{-3.0ex} \raisebox{2.2ex}{\tiny (-- \hspace*{-3.1ex} --)}
\rightarrow \dmkspip  \right) =
{A\bc_{i}^2 + A\bu_{i}^{2} \over 2}\ e ^{- {t\over \tau} }
\left\{ 1 \stackrel{(-)}{+} \right. \\
\hfill \displaystyle \left. {\cal R}_i\ {\mathrm {cos}}\ x{t \over \tau}
\stackrel{(-)}{+}
{\cal D}_i\ {\mathrm {sin}} \left(2\beta +\gamma +\Delta\delta_i \right)
{\mathrm {sin}} \ x{t \over \tau} \right\}
\label{eq:btodm}
\end{array}
\end{equation}
where $x = \Delta m \tau$, ${\cal R}_i= {\rho^2_i - 1 \over \rho^2_i + 1}\ ,\
{\cal D}_i = {2\rho_i \over \rho^2_i + 1} = \sqrt{ 1 - {\cal R}^2_i} $ with
$\rho_i$ = $ {A\bc_{i} \over A\bu_{i}}$ and $\Delta\delta_i = \delta\bc_{i} -
\delta\bu_{i}$. We have used the approximation $ \vert p/q \vert =1$.\\[-0.8ex]
The PDF of the transition 
$B^0 \hspace*{-3.0ex} \raisebox{2.2ex}{\tiny (-- \hspace*{-3.1ex} --)}
\rightarrow D^+ \ks \pi^-$ is
\begin{equation}
\begin{array}{l}
\displaystyle {\cal P}r \left(
B^0 \hspace*{-3.0ex} \raisebox{2.2ex}{\tiny (-- \hspace*{-3.1ex} --)}
\rightarrow D^+ \ks \pi^- \right) =
{A\bc_{i}^2 + A\bu_{i}^{2} \over 2}\ e ^{- {t\over \tau} }
\left\{ 1 \stackrel{(+)}{-} \right. \\
\hfill \displaystyle \left. {\cal R}_i\ {\mathrm {cos}}\ x{t \over \tau}
\stackrel{(-)}{+}
{\cal D}_i\ {\mathrm {sin}} \left(2\beta +\gamma -\Delta\delta_i \right)
{\mathrm {sin}} \ x{t \over \tau} \right\}
\label{eq:btodp}
\end{array}
\end{equation}
From the total number of events in each of the four final states
(\ref{eq:btodm}) and (\ref{eq:btodp}) and a global fit of their time
dependence, it is possible to extract the quantities, $ A\bc_{i}^2 +
A\bu_{i}^{2}\ ,\rho_i\ , S\ , \bar S$ where $S \equiv \sin(2\beta +\gamma
+\Delta\delta_i)$ and $\overline{S} \equiv \sin(2\beta +\gamma
-\Delta\delta_i)$. One obtains:
\vspace*{-1ex}
\begin{equation}
\sin^2(2\beta +\gamma ) = {1 \over 2} \left(1-S\overline{S} \pm 
\sqrt{(1-S^2)(1-\overline{S}^2)} \right)
\label{eq:sing_sq}
\end{equation}
Hence, in the limit of very high statistics, one would extract $\sin^2(2\beta
+\gamma )$ for each point $i$ of the Dalitz plot. However, for every point of
the Dalitz plot, $2\beta +\gamma$ is obtained with an eight-fold ambiguity in
the range $[0,2\pi]$, because of the invariance of the $\sin(2\beta +\gamma
\pm \Delta\delta_i )$ terms in Eq.~(\ref{eq:btodm}) and (\ref{eq:btodp})
under the three symmetry operations
\vspace*{-1ex}
\begin{equation}
\hspace*{-1ex}
\begin{array}{lll}
\sex   \hspace*{-2.3ex}&:~ 2\beta\!+\!\gamma \rightarrow \pi/2 - \Delta\delta,&
             \Delta\delta \rightarrow \pi/2 - (2\beta\!+\!\gamma )\\
\ssign \hspace*{-2.3ex}&:~ 2\beta\!+\!\gamma \rightarrow - (2\beta\!+\!\gamma ),&
             \Delta\delta \rightarrow \pi - \Delta\delta  \\
\spi   \hspace*{-2.3ex}&:~ 2\beta\!+\!\gamma \rightarrow 2\beta\!+\!\gamma + \pi,& 
	     \Delta\delta \rightarrow \Delta\delta + \pi
\end{array}
\hspace*{-2ex}
\label{eq:ambig}
\vspace*{-2ex}
\end{equation}
When the multiple measurements made in different points of the Dalitz plot
are combined, some of the ambiguity will be resolved, in the likely case that
the strong phase $\Delta\delta_i$ varies from one region of the Dalitz plot
to the other. This variation can either be due to the presence of resonances
or because of a varying phase in the non-resonant contribution.
In this case, the exchange symmetry $\sex$ is numerically different
from one point to the other, which in effect breaks this symmetry and
resolves the ambiguity.
Similarly, the $\ssign$ symmetry is broken if there exists some
{\it a-priori} knowledge of the dependence of $\Delta\delta_i$ on the Dalitz
plot parameters. This knowledge is provided by the existence of broad
resonances, whose Breit-Wigner phase variation is known and may be
assumed to dominate the phase variation over the width of the
resonance.
To illustrate this, let $i$ and $j$ be two points in the Dalitz plot,
corresponding to different values of invariant mass of a particular
resonance. One then measures $\sin(2\beta +\gamma \pm \Delta\delta_i)$ at
point $i$ and $\sin(2\beta +\gamma \pm (\Delta\delta_i + \alpha_{ij}))$ at
point $j$, where $\alpha_{ij}$ is known from the parameters of the
resonance.
It is important to note that the sign of $\alpha_{ij}$ is also known,
hence it does not change under $\ssign$. Therefore, should one choose
the $\ssign$-related solution $\sin(\pi -(2\beta +\gamma )\mp \Delta\delta_i
)$ at point $i$, one would get $\sin(\pi -(2\beta +\gamma) \mp
(\Delta\delta_i - \alpha_{ij}))$ at point $j$. Since this is different from 
$\sin(2\beta +\gamma \pm (\Delta\delta_i + \alpha_{ij}))$ the $\ssign$
ambiguity is resolved.
This is illustrated graphically in Eq.~\ref{eq:ssign}, where $\phi = 2\beta
+\gamma$ is the weak phase:
\vspace*{-2ex}
\begin{equation}
\hspace*{-1ex}
\begin{array}{ccc}
\sin(\phi \pm \Delta\delta_i) & \!\stackrel{\ssign}{\longleftrightarrow}\!&
\sin(\pi - \phi \mp \Delta\delta_i) \\[1ex]
BW \downarrow & & \downarrow BW \\[-0.5ex]
\sin(\phi \pm (\Delta\delta_i \!+\! \alpha_{ij})) 
& \stackrel{\ssign}{\not\!\!\!\longleftrightarrow}\! & 
\sin(\pi - \phi \mp (\Delta\delta_i \!-\! \alpha_{ij}) )
\end{array}
\hspace*{-1ex}
\label{eq:ssign}
\vspace*{-2ex}
\end{equation}
Thus, broad resonances reduce the initial eight-fold ambiguity to the
two-fold ambiguity of the $\spi$ symmetry, which is never broken.
Fortunately, $\spi$ leads to the well-separated solutions $2\beta +\gamma$ and
$2\beta +\gamma + \pi$, the correct one of which is easily identified when
this measurement is combined with other measurements of the unitarity
triangle.%

\section{The Finite Statistics Case}
\vspace*{-1ex}
Since experimental data sets will be finite, extracting $\gamma$ will
require making use of a limited set of parameters to describe the
variation of amplitudes and strong phases over the Dalitz plot. The
consistency of this approach can be verified by comparing the results
obtained from fits in a few different regions of the Dalitz plot, and the
systematic error due to the choice of the parameterization of the data may be
obtained by using different parameterizations.

A fairly general parameterization assumes the existence of $N_R$
Breit-Wigner resonances, as well as a non-resonant contribution:

\begin{equation}
\begin{array}{l}
{\cal A}\,_\xi(\bztodm) = \\
        \hfill \left(A\bc_{0}\, e^{i\delta\bc_{0}} + 
	\sum_{j=1}^{N_R} A\bc_{j}B_{s_j}(\xi) \, e^{i\delta\bc_{j}} \right)
	e^{i\delta\bc(\xi)}
	\nonumber\\
{\cal A}\,_\xi(\bztodp) = \\
        \hfill \left(A\bu_{0}\, e^{i\delta\bu_{0}} + 
	\sum_{j=1}^{N_R} A\bu_{j}B_{s_j}(\xi) \, e^{i\delta\bu_{j}} \right)
	e^{i (\delta\bu(\xi) + \gamma)}
\end{array}
\label{eq:amp-param}
\end{equation}
where $\xi$ represents the Dalitz plot variables, $B_{s_j}(\xi) \equiv
b_{s_j}(\xi)\, e^{i\delta_j(\xi)}$ is the Breit-Wigner amplitude for a
particle of spin $s_j$, normalized such that $\int |b_{s_j}(\xi)|^2 d\xi=1$,
$A\bu_0$ and $\delta\bu_0$ ($A\bc_0$ and $\delta\bc_0$) are the
magnitude and CP-conserving phase of the non-resonant $\bbartoubar$
($\bbartocbar$) amplitude, 
and $A\bu_j$ and $\delta\bu_j$ ($A\bc_j$ and $\delta\bc_j$) are the
magnitudes and CP-conserving phase of the $\bbartoubar$
($\bbartocbar$) amplitude associated with resonance $j$.
%
The functions $\delta\bc(\xi)$ and $\delta\bu(\xi)$ may be assumed to
vary slowly over the Dalitz plot, allowing their description in terms
of a small number of parameters. 
The decay amplitudes of $\Bbar$ mesons are identical to those of
Eqs.~(\ref{eq:amp-param}), with $\gamma$ replaced by $-\gamma$.

The decay amplitudes of Eqs.~(\ref{eq:amp-param}) can be used to conduct the
full data analysis. Given a sample of $N_e$ signal events, $\gamma$ and the
other unknown parameters of Eq.~(\ref{eq:amp-param}) are determined by
minimizing the negative log likelihood function
\vspace*{-1ex}
\begin{equation}
\chi^2 \equiv -2 \sum_{i=1}^{N_e} \log P(\xi_i), ~~~\mbox{with}~~~
P(\xi) = |{\cal A}\,_{\xi}(f)|^2,
\label{eq:chi2}
\vspace*{-1ex}
\end{equation}
where the amplitude $A\,_{\xi}(f)$ is given by one of the expressions of
Eqs.~(\ref{eq:amp-param}), or their CP-conjugates, depending on the final
state $f$, and $\xi_i$ are the Dalitz plot variables of event $i$.

In what follows, we discuss important properties of the method by
considering the illustrative case, in which the $\bbartoubar$ decay
proceeds only via a non-resonant amplitude, and the $\bbartocbar$
decay has a non-resonant contribution and a single resonant
amplitude. For concreteness, the resonance is taken to be the
$K^{*\pm}(892)$. We take the a priori $\xi$-dependent non-resonant
phases to be $\delta\bc(\xi) = \delta\bu(\xi) = 0$.

\section{Simulation Studies}
\vspace*{-1ex}
\label{sec:simulation}
To study the feasibility of the analysis using Eq.~(\ref{eq:chi2}), we
conducted a simulation of the decays $\bztodmp$ and $\bzbtodpm$.
Events were generated according to the PDF in Eqs.~(\ref{eq:btodm}) and
~(\ref{eq:btodp}), with the parameter values given in
Table~\ref{tab:reference}.
In this table and throughout the rest of the paper, we use a tilde to
denote the ``true'' parameter values used to generate events, while
the corresponding plain symbols represent the ``trial'' parameters
used to calculate the experimental $\chi^2$.
For simplicity, additional resonances were not included in this
demonstration. However, broad resonances that are observed in the
data should be included in the actual data analysis.
\begin{table}[ht!]
\begin{center}
\begin{tabular}{cc|cc}
  Parameter              &Value   &Parameter                           &Value \\
\hline
$\tilde{\delta}\bc(\xi)$ &0       &$\tilde A\bu_0 / \tilde A\bc_0$     &0.4\\
$\tilde{\delta}\bu(\xi)$ &0       &$\tilde A\bc_{K^*} / \tilde A\bc_0$ &1.0\\
$\tilde\delta_{K^*}$     &1.8     &$\tilde A\bc_{K^*}$                 &$\sqrt{2 \times 10^{-4} \Gamma_B}$\\
$\tilde\delta\bu_0$      &1.0     &$\widetilde{2\beta\!+\!\gamma}$     &2.00\\
\hline
\end{tabular}
\end{center}
\vspace*{-3ex}
\caption{Parameters used in the simulation.
	The value of $\tilde A\bc_{K^*}$ is chosen so as to 
	roughly agree with the measurement of the corresponding branching 
	fraction~\protect\cite{ref:cleo-btodkst}.}
\label{tab:reference}
\end{table}

The simulations were conducted with a benchmark integrated luminosity
of 400~\fb.
The reconstruction efficiencies were based on
current $\Upsilon(4{\rm S})$ detector capabilities. We assumed
an efficiency of 50\% for reconstructing the $K_s$, and 90\% for
reconstructing the $\pi^\pm$. The product of reconstruction efficiencies
and branching fractions of the $D^\pm$, summed over the final states
$K^\mp\pi^\pm\pi^\pm$ and $K_s\pi^\pm$ is taken to be 4\%. Finally the
analysis efficiency {\it including perfect tagging} was estimated to be
20\%.
The numbers of signal events obtained in each of the final states with
the above efficiencies and the parameters of Table~\ref{tab:reference}
are listed in Table~\ref{tab:nsignal}.

\begin{figure}[htbp!]
\hbox to\hsize{\hss
\includegraphics[width=\hsize]{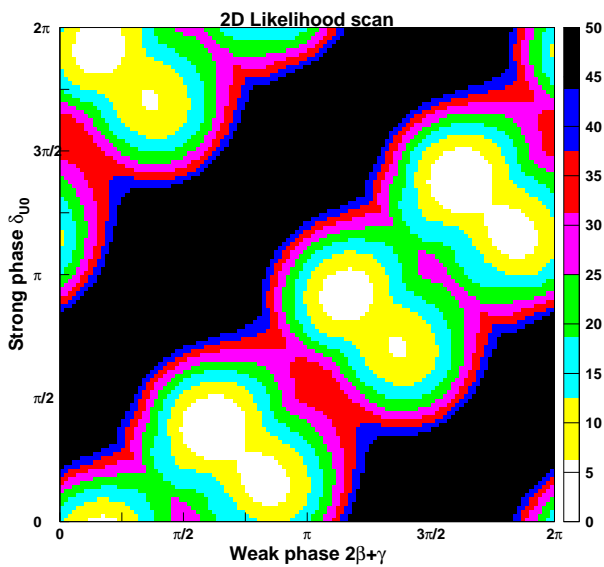}
\hss}
\vspace*{-3ex}
\caption{$\chi^2$ as a function of $2\beta +\gamma$ and
$\delta\bu_0$, with the parameters of Table~\ref{tab:reference}.}
\label{fig:scan-nr}
\end{figure}

In Fig.~\ref{fig:scan-nr} we show the $\chi^2$ dependence as a function of
$2\beta +\gamma$ and $\delta\bu_0$. The smallest value of $\chi^2$ is shown
as zero (white).
At each point in these figures, the $\chi^2$ is calculated with the
generated values of the amplitude ratios $A\bu_0 / A\bc_0 = \tilde
A\bu_0 / \tilde A\bc_0$ and $A\bc_{K^*}/ A\bc_0 = \tilde
A\bc_{K^*}/ \tilde A\bc_0$.  
When these amplitude ratios are determined by a fit simultaneously
with the phases, the correlations between the amplitudes and the
phases are found to be less than 10\%.
Therefore, the results obtained with the amplitudes fixed to their
true values are sufficiently realistic for the purpose of this
demonstration.

\vspace*{-4ex}

\begin{figure}[htbp!]
\vspace*{-3ex}
\hbox to\hsize{\hss
\includegraphics[width=\hsize]{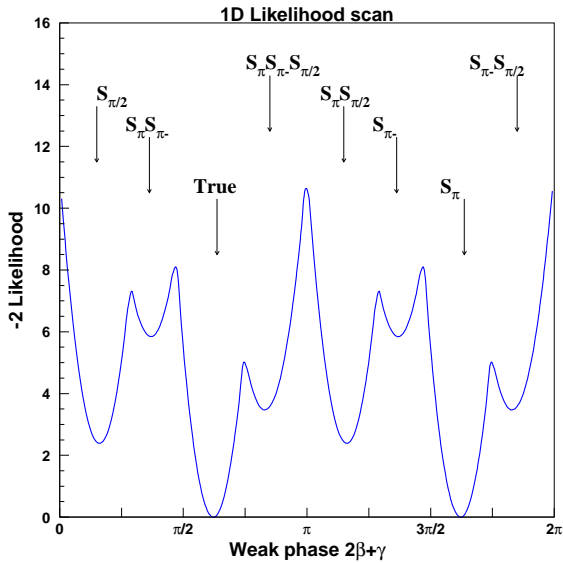}
\hss}
\vspace*{-4ex}
\caption{Minimum projection of $\chi^2$ onto $2\beta +\gamma$.}
\label{fig:scan-nr-1d}
\end{figure}

In Fig. \ref{fig:scan-nr-1d} we show the one-dimensional minimum projection
$\chi^2(2\beta +\gamma)= \min\{ \chi^2(2\beta +\gamma, \delta\bu_0)\}$,
showing the smallest value of $\chi^2$ for each value of $2\beta +\gamma$.

With equal resonant and non-resonant $\bbartocbar$ amplitudes, the
ambiguities originating from the non-resonant regime are dominant, due to the
relative suppression of the resonant interference terms, as they overlap only
little with the non-resonant amplitudes. These are however significant enough
to resolve all but the $\spi$ ambiguity.

\vspace*{-5ex}
\begin{figure}[htbp!]
\hbox to\hsize{\hss
\includegraphics[width=1.1\hsize]{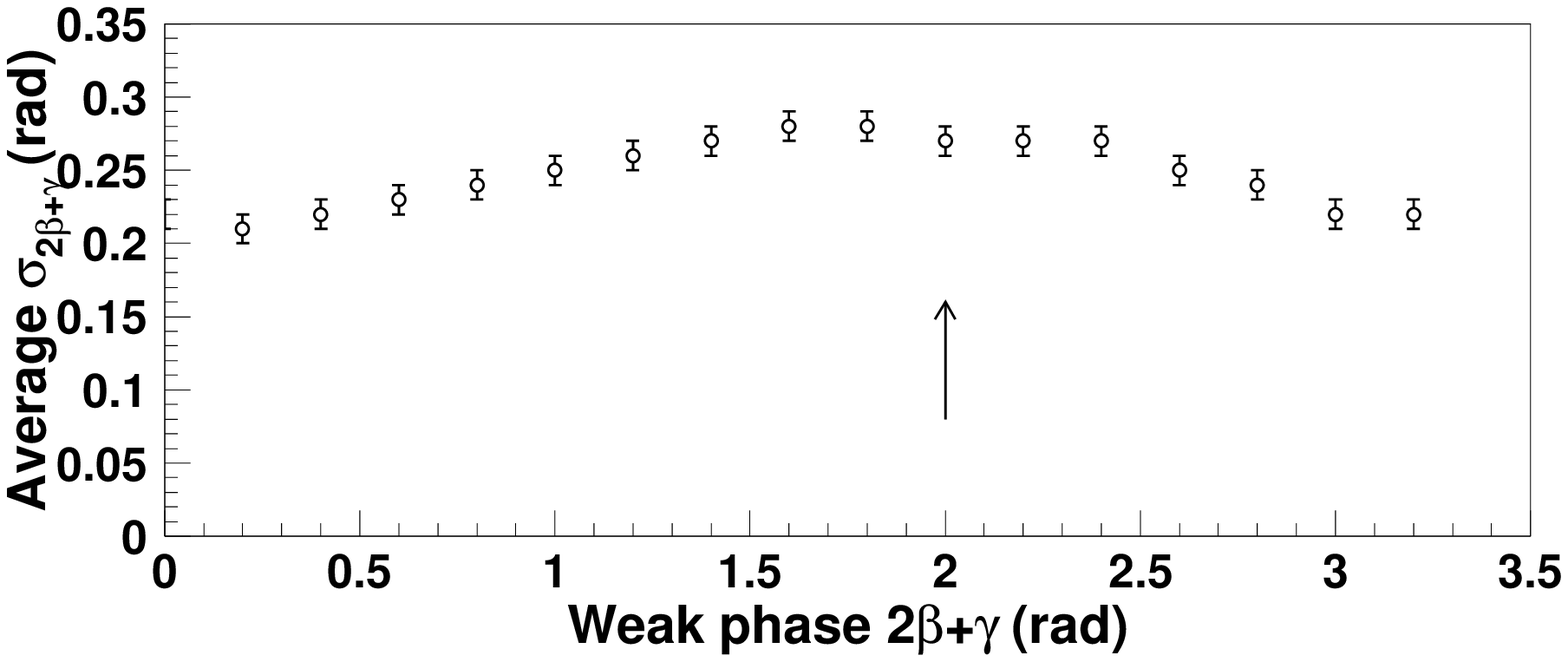}
\hss}
\vspace*{-8ex}
\end{figure}
\vspace*{-5ex}
\begin{figure}[htbp!]
\hbox to\hsize{\hss
\includegraphics[width=1.1\hsize]{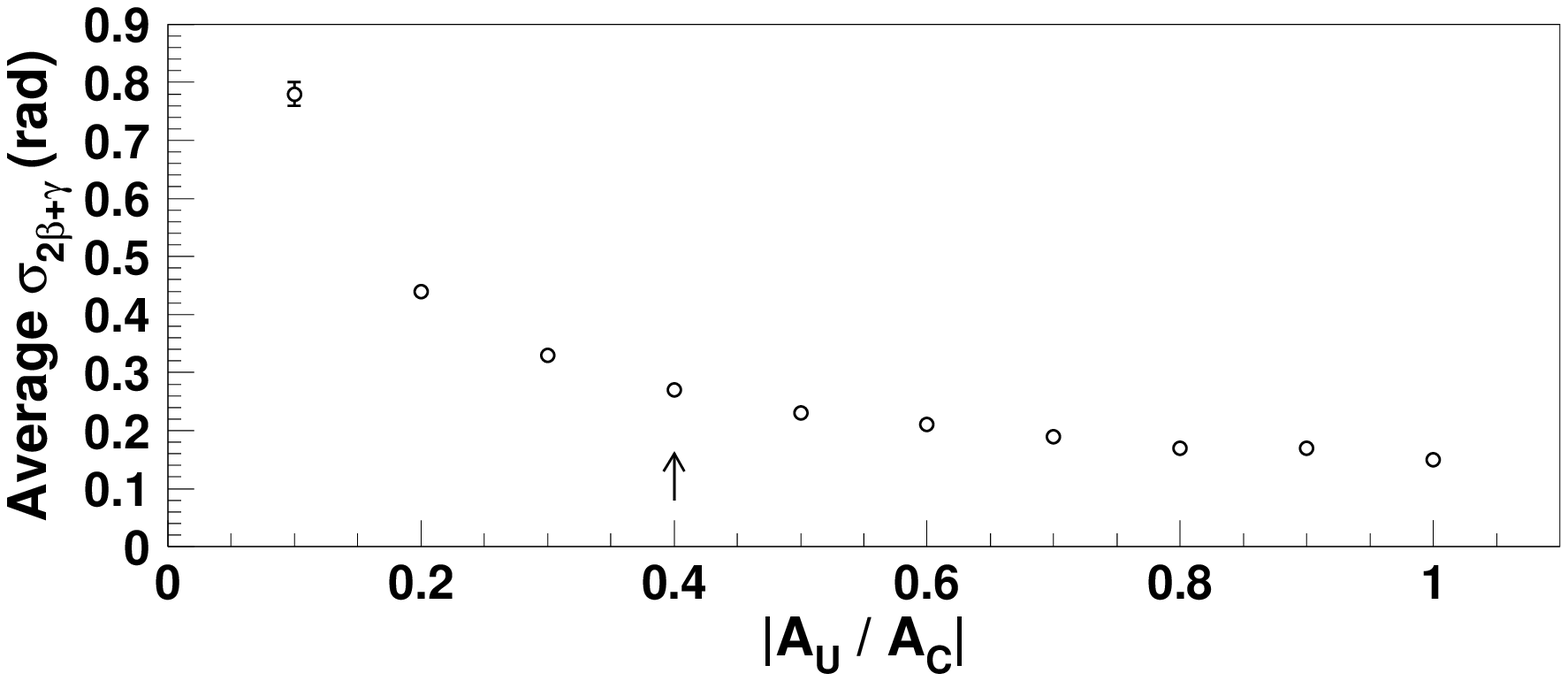}
\hss}
\vspace*{-4ex}
\caption{The error on $2\beta +\gamma$, $\sigma_{2\beta +\gamma}$, as a
function of ({\bf top}) $\widetilde{2\beta\!+\!\gamma}$ and ({\bf bottom})
$\tilde A\bu_0 / \tilde A\bc_0$.}
\label{fig:sig-gamma-vs-gamma}
\end{figure}

\begin{table}[hbtp!]
\begin{center}
\begin{tabular}{lr|lr}
Mode       &$N_{ev}$ &Mode       &$N_{ev}$\\
\hline
$\bztodm$  & 112     &$\bztodp$  &  33 \\ 
$\bzbtodp$ & 111     &$\bzbtodm$ &  33 \\
\hline
\end{tabular}
\end{center}
\vspace*{-3ex}
\caption{The numbers of perfectly tagged events of each type expected in 400
\fb\ obtained by averaging 100 simulations using the parameters of
Table~\protect\ref{tab:reference} and the reconstruction efficiencies listed
in the text.}
\label{tab:nsignal}
\end{table}

In Fig.~\ref{fig:sig-gamma-vs-gamma} we present $\sigma_{2\beta +\gamma}$,
the statistical error on $2\beta +\gamma$ as a function of
$\widetilde{2\beta\!+\!\gamma}$ and $\tilde A\bu_0 / \tilde A\bc_0$, obtained
by fitting simulated event samples.
Each point is obtained by repeating the simulation 
250 times, to minimize sample-to-sample statistical fluctuations. All the
parameters of Table~\ref{tab:reference} were determined by the fit, and the
arrow in these figures indicates the point corresponding to the parameters of
Table~\ref{tab:reference}.
One observes that $\sigma_{2\beta +\gamma}$ does not depend strongly on the
value $2\beta +\gamma$. As expected, a strong dependence on $\tilde A\bu_0 /
\tilde A\bc_0$ is seen. However, it should be noted that $\sigma_{2\beta
+\gamma}$ changes very little for all values of $\tilde A\bu_0 / \tilde
A\bc_0$ above 0.4. This suggests that the potential for a significantly
sensitive measurement is high over a broad range of parameters.

\section{Conclusions}
\vspace*{-1ex}
We have shown how $2\beta +\gamma$ may be measured in the color-allowed decays
\btodkpiall, focusing on the simplest mode \bdkpi. The absence
of color suppression in the $\bbartoubar$ amplitudes is expected to
result in relatively large rates and significant CP violation effects,
and hence favorable experimental sensitivities. While the Dalitz plot
analysis implies some experimental complication, it should reduce the
eight-fold ambiguities, which are a serious problem with other methods for
measuring $2\beta +\gamma$.
As a result of these advantages, this method is likely to lead to
a measurement of $2\beta +\gamma$ even with the current generation of
$B$~factory experiments.

\section{Acknowledgments}
\vspace*{-1ex}
The authors thank Francois Le Diberder for his fruitful ideas and help with
simulation.
This work was supported by CEA and the Danish Research Academy.

\end{document}